\documentclass[conference]{IEEEtran}
\IEEEoverridecommandlockouts

\usepackage{cite}
\usepackage{amsmath,amssymb,amsfonts}
\usepackage{algorithmic}
\usepackage{graphicx}
\usepackage{textcomp}
\usepackage{xcolor}
\usepackage{url}

\def\BibTeX{{\rm B\kern-.05em{\sc i\kern-.025em b}\kern-.08em
    T\kern-.1667em\lower.7ex\hbox{E}\kern-.125emX}}
\begin{document}

\title{BLONDiE: Blockchain Ontology with Dynamic Extensibility\\
\thanks{978-1-7281-5628-6/20/\$31.00 ©2020 IEEE}
}

\author{\IEEEauthorblockN{1\textsuperscript{st} H\'ector Eduardo Ugarte-Rojas}
\IEEEauthorblockA{\textit{Department of Informatics} \\
\textit{Universidad Nacional de San Antonio Abad del Cusco}\\
Cusco, Peru \\
hector.ugarte@unsaac.edu.pe}
\and
\IEEEauthorblockN{2\textsuperscript{nd} Boris Chullo-Llave}
\IEEEauthorblockA{\textit{Department of Informatics} \\
\textit{Universidad Nacional de San Antonio Abad del Cusco}\\
Cusco, Peru \\
boris.chullo@unsaac.edu.pe}
}

\maketitle

\begin{abstract}
There are thousands of projects worldwide based primarily on blockchain technology. 
These have a large number of users and hundreds of use cases. One of the most popular
is the use of cryptocurrencies and their benefits against money without intrinsic value 
(fiat money) and centralized financial solutions. However, although thousands of new 
transactions are carried out daily in different platforms, uniform and standardized 
information does not exist to be able to manage the large amount of data that is 
generated and exchanged between users through transactions and the generation of new blocks. 

This research reports the development of BLONDiE, an ontology that allows the semantic 
representation of knowledge to describe the native structure and related information of 
the three most relevant blockchain projects to date: Bitcoin, Ethereum and in the recent 
1.0 version extends its definitions to include Hyperledger, specifically the Hyperledger 
Fabric infrastructure. Its use allows having common data formats of different platforms 
for further processing, such as the execution of semantic queries.
\end{abstract}

\begin{IEEEkeywords}
Ontologies, Blockchain, RDF, OWL, Semantic Web
\end{IEEEkeywords}

\section{Introduction}
\begin{figure}[h!]
    \centering
    \includegraphics[scale=0.23]{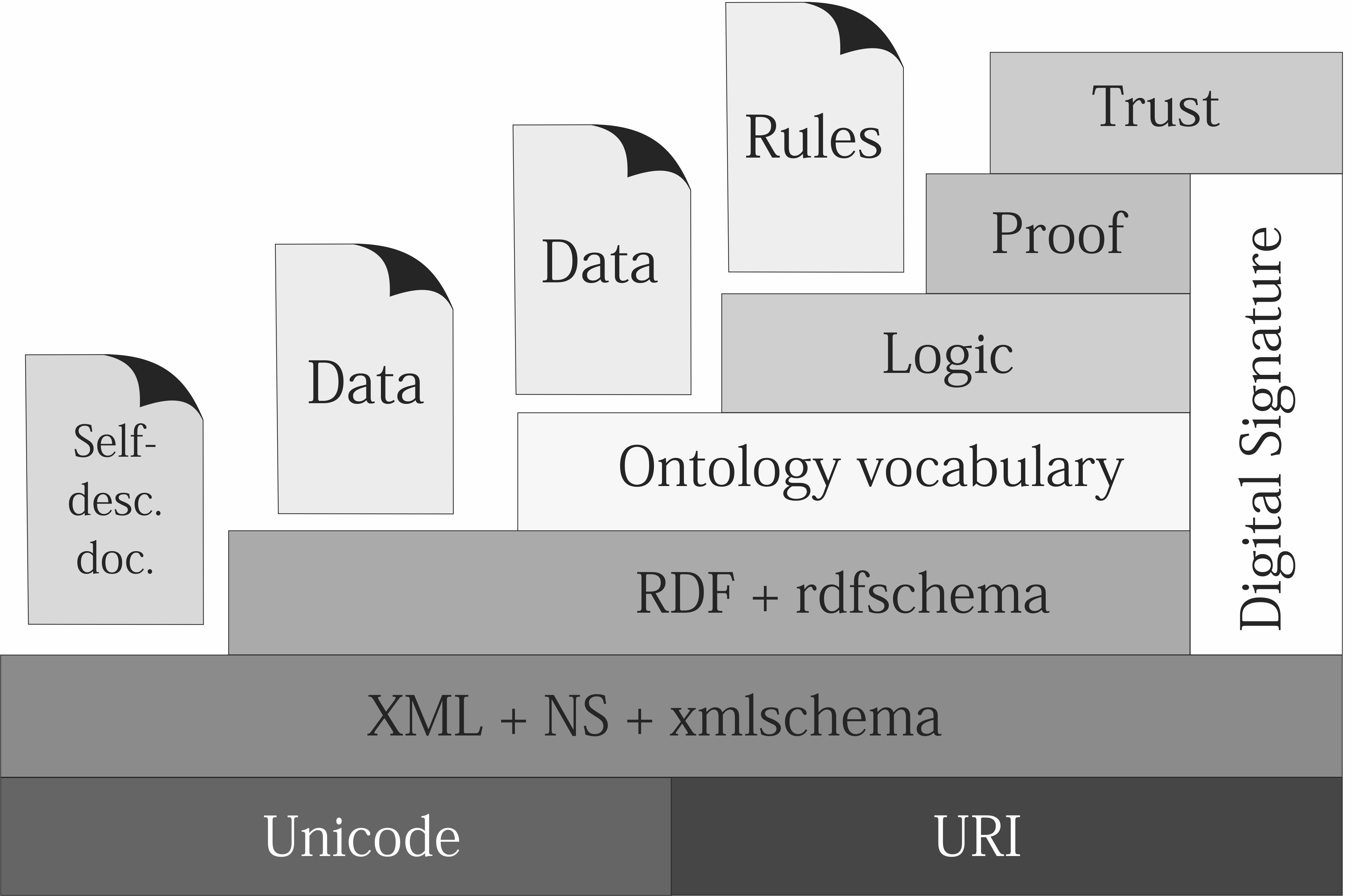}
    \caption{Stack of Semantic Web technologies \cite{berners2000semantic}}
    \label{fig:1}
\end{figure}

In computer science, ontologies are used as tools to represent, name, and define categories,
properties and relationships between concepts, data and entities of one or more 
domains~\cite{1BHLM2015}. These ontologies are part of the stack of semantic web 
technologies proposed by Tim Berners-Lee as seen in Figure~\ref{fig:1}. The semantic 
web, originally intended for the WWW (World Wide Web) provides a common framework for 
sharing and reuse data in different applications and companies~\cite{2w3c}. The vast 
majority of WWW data is readable for humans, but not for computers. The semantic web 
allows changing this reality. Once you have data into a readable form for computers 
it is possible to generate intelligent agents that can easily relate resources of 
different natures.

Blockchain is a paradigm born thanks to the anonymous implementation of Bitcoin, 
a digital currency, decentralized and based on cryptography~\cite{3Nakamoto2008}. 
With its creation in 2008, it was possible for the first time to maintain a 
non-centralized database, ensuring the immutability of the data.
A few years later, the use of the protocol for the management of other digital 
assets began to be studied. A large number of new projects emerged and also more general-purpose platforms such as Ethereum~\cite{4Wood2014} 
and Hyperledger~\cite{5} that permit coding blockchain 
applications in what is known as ``Smart Contracts''.

A smart contract is a tool to automate agreements between two or more parties. 
It is an algorithm that can be self-executed, self-fulfilled, 
self-verified and self-restricted according to the rules of how it has been 
coded~\cite{7Swanson2014}. The term coiner Nick Szabo states that ``Smart 
contracts combine protocols, users interfaces, and promises expressed via those 
interfaces, to formalize and secure relationships over public networks''~\cite{8Szabo1997}.

With a smart contract, it is possible to generalize the idea of Bitcoin for multiple 
purposes. For example, to register land titles or manage fruits as apples from harvest 
by the farmer to sale in a supermarket. In the same way that a user guarantees 
the possession of 1 bitcoin, it is possible to demonstrate the possession of a parcel or an apple in a given time.

With the previous example, the importance of the blockchain paradigm for supply 
chain management is clear. A producer generates a new token representing a physical 
entity, he transfers it to a distributor. Such company to a retailer, and finally 
the retailer to an end-user or customer. Being an immutable database guarantees 
the veracity of the data. Using public-key cryptography ensures that only the 
actor with the right private key can sign new transactions~\cite{9Hector2016}.

Then with so many worldwide nodes running different blockchains and 
many of them have sensitive and highly important records, It is fundamental 
to have standardized, structured and homogeneous data. That, among other things, 
allows the use of ontologies.

There is no standardized methodology for the development of ontologies, much 
less a single methodology that is correct~\cite{10NDMc2001}. In addition, a 
single ontology cannot cover a whole domain of knowledge, so the ontologies are 
made to be extensible and reusable.

This article details the development of BLONDiE (\textbf{BL}ockchain \textbf{ON}tology with \textbf{D}ynamic \textbf{E}xtensibility) , an ontology expressed in OWL language (Web Ontology Language) 
called that way to be easily remembered, modified, extended, and improved by new definitions or other ontologies over time (what we consider as dynamism). This allows the integration of native structural data and related information from uneven sources of the three relevant blockchain projects: Bitcoin, Ethereum and Hyperledger Fabric. Formally defined semantics will make it possible to perform precise searches and complex queries with the SPARQL (SPARQL Protocol and RDF Query Language) language. We believe that BLONDiE is an innovative ontology since there is no relevant prior work that accurately covers its scope and domain.

The use of ontologies in blockchain technologies is the first step in the realization of the ``Semantic Blockchain''. An emerging paradigm that considers the existence of 
new networks where the participants have the certainty that the meaning of the messages is the same for all and there is trust about the arranged circumstances and 
agreements~\cite{9Hector2016}. 

\section{Background} 
\subsection{BLOCKCHAIN} 
\begin{figure}[h!]
    \centering
    \includegraphics[scale=1]{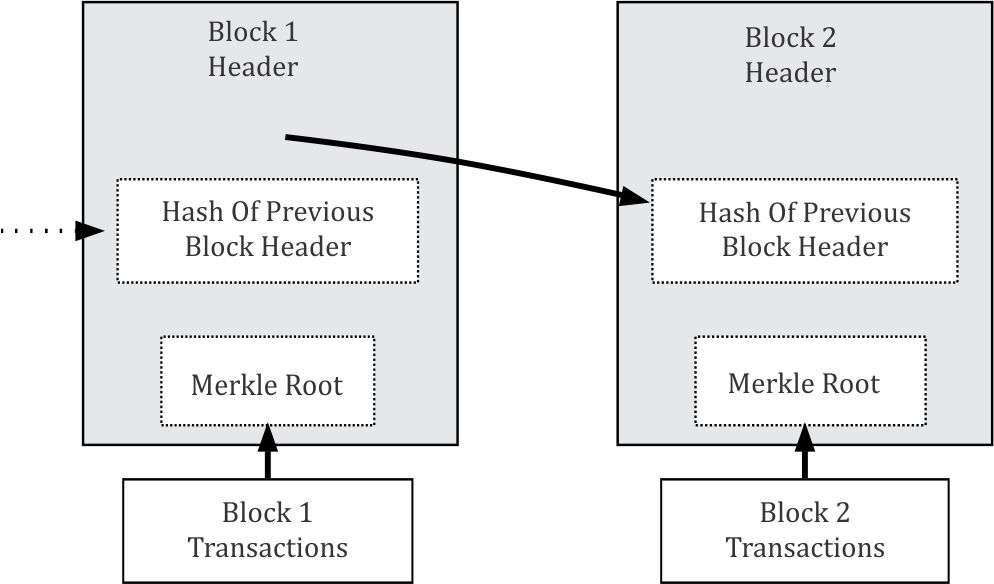}
    \caption{Simplified structure of the Bitcoin blockchain\cite{bitcoin2015}}
    \label{fig:2}
\end{figure}

It is a data structure and a database, a record of all confirmed and processed transactions that occurred in a particular network (Bitcoin, Ethereum, etc.) since its inception. It is shared and replicated by all participating nodes thanks to the P2P (Peer to Peer) architecture. These transactions are stored in blocks that share the same temporary origin, future transactions will form new blocks. Hence the given name emulating a chain of blocks, from the zero block to the current block (see Figure~\ref{fig:2})~\cite{Antono2014}.

The generation time of a new block depends on the protocol. For example, in Bitcoin is approximately 10 minutes~\cite{3Nakamoto2008}, but in Ethereum is 15 seconds~\cite{4Wood2014}. The integrity of the data at different levels is guaranteed thanks to cryptography, each participant has a private key to sign transactions, also each block is linked to the previous one by storing the hash of the previous block.

Special nodes called miners are responsible for producing new blocks. On public networks, they receive cryptocurrency tokens as payment for their time spent solving a mathematical 
problem called proof of work. In private networks such as Hyperledger, this bonus is not necessary since it is restricted who is allowed to participate in it.

\subsubsection{Bitcoin}
It is the world's first decentralized digital currency created by an anonymous under the pseudonym of Satoshi Nakamoto. It is the first time that the blockchain architecture is used formally. 1 bitcoin is divisible up to 100,000 pieces called satoshis.

\subsubsection{Ethereum}
It was proposed in 2013 by Vitalik Buterin. It is a platform similar to Bitcoin cryptocurrency, but offers a Turing complete virtual machine (i.e., you can solve all computational problems) through smart contracts.

\subsubsection{Hyperledger Fabric}
Hyperleger is an umbrella project (union of multiple interested entities) started by the Linux Foundation to advance cross-industry collaboration on blockchain technology. One of those projects is Hyperledger Fabric. 
This is contributed by IBM and allows its components to be ``plug and play''.

\subsection{RDF}

From its acronym: ``Resource Description Framework`'. It is a language and metadata model recommended by the W3C (The World Wide Web Consortium) to encode knowledge and build a readable semantic infrastructure for computers and electronic agents seeking information 
on the Web~\cite{miller1998introduction}. The modeled resources must have a ``Universal Resource Identifier`' (URI). To describe resources you have a set of properties. Descriptions are statements in the subject-predicate-object structure and commonly called triplets, defining a graph structure~\cite{Gutierrez2005}.

A basic triplet is: Bitcoin is a cryptocurrency.

\subsection{OWL}
OWL (Web Ontology Language) is a language made to represent complex and varied knowledge about things, sets of things, and relationships between them. It is a language for ontologies more expressive than XML, RDF, and RDFS that provides additional vocabulary and more formal semantics~\cite{14Cardoso2015}. It has a diverse set of operations such as union, intersection, complement, etc.

\subsection{SPARQL}
SPARQL (recursive acronym for “SPARQL Protocol and RDF Query Language”). It is a language recommended by the W3C for semantic queries for data sets, made to handle and return data stored in RDF format. Therefore, the queries work on the structure of a graph defined by RDF data, where the result will also be a graph or subset of it.

\section{Ontology development methodology}
For BLONDiE development we rely on the iterative methodology proposed by Noy and McGuiness~\cite{10NDMc2001}, and summarized as:

\begin{itemize}
    \item Define the domain and scope.             
    \item Define classes and the class hierarchy.            
    \item Define the properties of the classes.             
    \item Define the facets of the slots.             
    \item Create instances.             
\end{itemize}

Thanks to its practicality, simplicity, and ease of application, we consider this ontology as the most appropriate in comparison to others.

\textbf{Domain and scope:}
The ontology developed covers the domain of blockchains in three relevant blockchain technologies: Bitcoin, Ethereum and Hyperledger Fabric. It is a specific domain ontology, where the concepts with relationships and properties are a new means of storing and propagating specialized knowledge.

The scope is the description of the native structure of these data structures and related information. From here we state that it is an unprecedented ontology, after BLONDiE another ontology called ethOn~\footnote{\url{https://consensys.github.io/EthOn/EthOn_spec.html}} was developed, but it is limited only to the Ethereum network.

There are other ontologies with another scope and focus, for example de Kruijff and Weigand~\cite{Towards2017} developed and ontology focused on business operations and processes of potential enterprises making a distinction between the datalogical, infological, and essential level of blockchain transactions and smart contracts.

\begin{figure*}
    \centering
    \includegraphics[scale = 0.6]{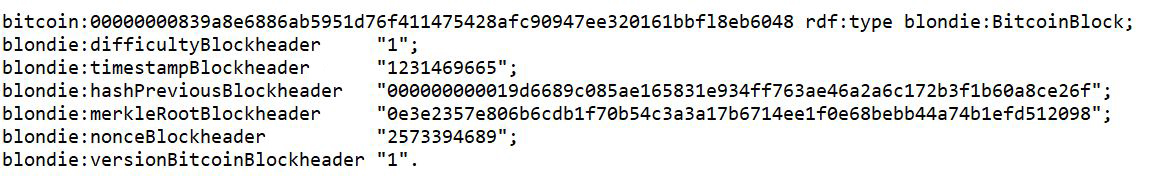}
    \caption{Bitcoin block described in RDF using BLONDiE.}
    \label{fig:3}
\end{figure*}
\begin{figure*}
    \centering
    \includegraphics[scale = 0.55]{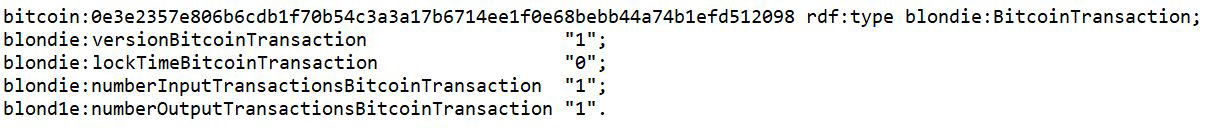}
    \caption{SPARQL query using BLONDiE}
    \label{fig:4}
\end{figure*}

\begin{figure}
    \centering
    \includegraphics[scale = 1]{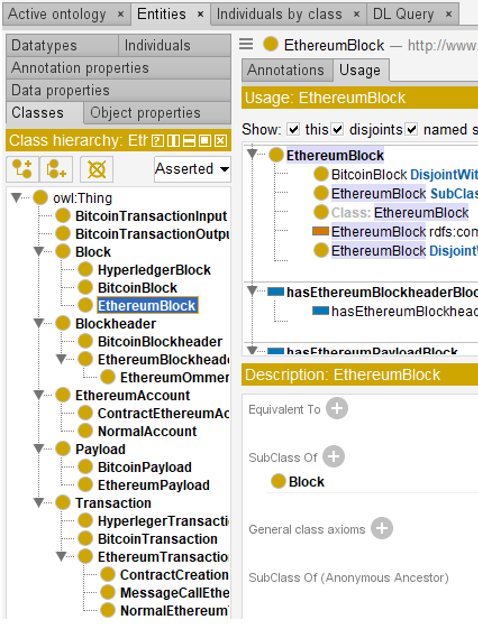}
      \caption{Protege graphic interface v.5.5.0.}
    \label{fig:5}
\end{figure}

The chosen competency questions are:
\begin{itemize}
    \item \textbf{CQ1}. Who was the miner of each block?
    \item \textbf{CQ2}. What is the height of each block?
    \item \textbf{CQ3}. How many transactions were included in a block?
    \item \textbf{CQ4}. Is a transaction confirmed or unconfirmed?
    \item \textbf{CQ5}. How many coins in total were transferred in a block?
\end{itemize}

\begin{figure*}
    \centering
    \includegraphics[scale=0.19]{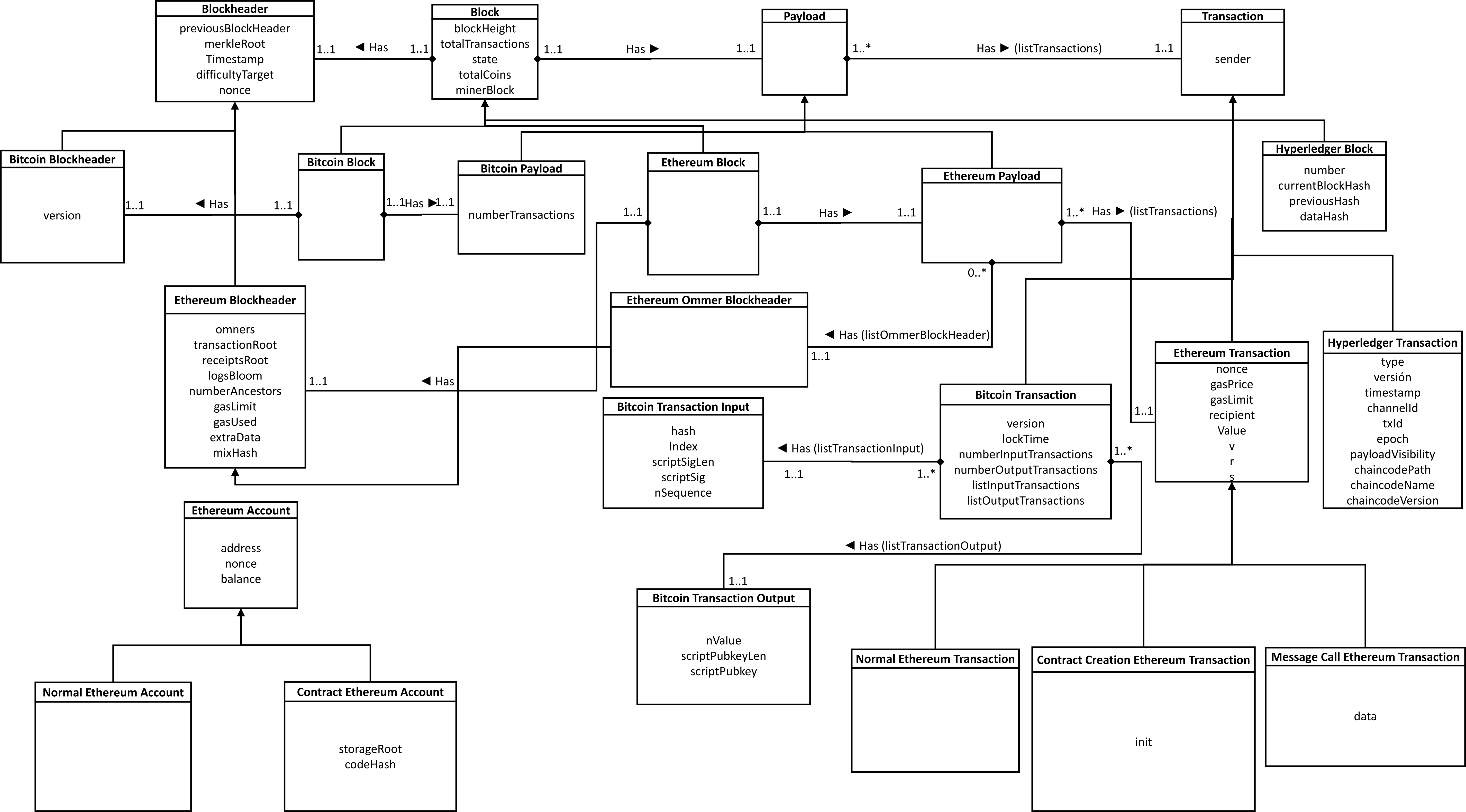}
    \caption{BLONDiE Extended Entity Relationship Diagram}
    \label{fig:6}
\end{figure*}

\subsection{CLASSES, PROPERTIES AND FACETS}
To determine the classes, properties and facets we will rely on the documented native structure of these technologies. These are detailed as follows: Table~\ref{tab:1} and~\ref{tab:2} for bitcoin, table~\ref{tab:3},~\ref{tab:4},~\ref{tab:5} and~\ref{tab:6} for Ethereum and table~\ref{tab:7} and~\ref{tab:8} for Hyperledger Fabric.

\begin{table}[h!]
    \centering

    \caption{Structure of a Bitcoin block~\cite{Antono2014, okupski2014bitcoin}}
        \label{tab:1}
    \begin{tabular}{p{1.5cm}cp{3.75cm}}\hline
       \textbf{Field Name} &  \textbf{Type (Size)} &  \textbf{Description} \\\hline\hline
        nVersion & int (4 Bytes) & A version number to track software/protocol upgrades. \\
        HashPrevBlock & uint256 (32 bytes) & A reference to the hash of the previous (parent) block in the chain                                             $SHA256^2$. (nVersion~$| \dots |$ nNonce).\\
        HashMerkleRoot & uint256 (32 bytes) & A hash of the root of the merkle tree of this block’s transactions.\\
        nTime &  unsigned int (4 bytes) & The approximate creation time of this block (seconds  from Unix Epoch)\\
        nBits & unsigned int (4 bytes) & The size of the block, in bytes, following this field.\\
        nNonce & unsigned int (4 bytes) & A counter used for the proof-of-work algorithm. \\
        \#vtx (Transaction counter) & VarInt (1-9 bytes) & How many transactions follow in vtx .\\
        vtx[] & Transaction (Variable) & Transaction recorded in this block.\\\hline
    \end{tabular}
\end{table}

\begin{table}[h!]
    \centering
        \caption{Structure of a Bitcoin transaction~\cite{Antono2014,okupski2014bitcoin}}
        \label{tab:2}
       \begin{tabular}{p{0.5cm}p{1.5cm}p{1.5cm}p{3.5cm}}\hline
       \multicolumn{2}{c}{\textbf{Field Name}}  &  \textbf{Type (Size)} &  \textbf{Description} \\\hline\hline
        nVersion  & & Int (4 bytes) & Transaction version format (Currently 1).\\
        \#vin     & &  VarInt (1-9 bytes) & Number of transaction entries. \\
        vin[]     & hash  & uint256 (32 bytes) &  SHA256 Double hash of a past transaction.\\
                  & n  & uint (4 bytes) &  Index of an outbound transaction in the specific hash transaction. \\
                  & scriptSigLen &  VarInt (1-9 bytes) & Length of the scriptSign field in bytes \\
                  & scriptSig & CScript (Variable) & Script that satisfies the spent condition of the outbound transaction (hash, n).\\
                  & nSequence & uint (4 bytes) & Sequence number of the incoming transaction. \\
        \#vout    & & VarInt (1-9 bytes) & Number of outbound transaction entries in vout.\\
        vout[]    & nvalue & int64 t (8 bytes) & Amount of $10^{-8}$ BTC.\\
                  & scriptPubkeyLen & VarInt (1-9 bytes) & Length of the scriptPubkey field in bytes. \\
                  & scriptPubkey & CScript (Variable) & Script that specifies conditions in which the transaction exit can be claimed.\\
        nLockTime & & Unsigned int (4 bytes) & Past timestamp in which transactions can be replaced before the inclusion of a block. \\
        \\\hline
    \end{tabular}
\end{table}

\begin{table}[h!]
    \centering
        \caption{Structure of an Ethereum block~\cite{4Wood2014}.}
    \label{tab:3}
    \begin{tabular}{p{2.2cm}p{6cm}}\hline
       \textbf{Field Name}  &  \textbf{Description} \\\hline\hline
         parentHash       & The SHA3 256-bit hash of the parent block’s header, in its entirety.\\
         
         ommersHash       & The SHA3 256-bit hash of the ommers list portion of this block. \\
         beneficiary      & The 160-bit address to which all fees  collected from the successful mining of this block
                            be transferred. \\
         stateRoot        & The  SHA3 256-bit hash of the root  node of the state trie, after all transactions are
                              executed and finalisations applied. \\
         transactionsRoot & The SHA3 256-bit hash of the  root node of the trie structure populated with each 
                            transaction in the transactions list portion  of the block. \\
         receiptsRoot     & The SHA3 256-bit hash of the root  node of the trie structure populated with
                            the receipts of each transaction in the transactions list  portion of the block.\\
         logsBloom        & The Bloom filter composed from indexable information (logger address and log 
                            topics) contained in each log entry from the receipt of each transaction in the transactions list.\\
         difficulty       & A scalar value corresponding to the difficulty level of this block. This can be calculated
                            from the previous block’s difficulty level and the timestamp. \\
         number           & A scalar value equal to the number of ancestor blocks. The genesis block has a number of zero. \\
         gasLimit         & A scalar value equal to the current limit  of gas expenditure per block. \\
         gasUsed          & A scalar value equal to the total gas used  in transactions in this block.\\
         timestamp        & A scalar value equal to the reasonable  output of Unix’s time() at this block’s inception. \\
         extraData        & An arbitrary byte array containing  data relevant to this block. This must be 32 bytes  or fewer. \\
         mixHash          &  A 256-bit hash which proves combined  with the nonce that a sufficient amount of 
                            computation has been carried out on this block. \\
         nonce            & A 64-bit hash which proves combined with  the mix-hash that a sufficient amount 
                            of computation has been carried out on this block.\\
         transactions[]   & List of transactions.\\
         ommersblockheaders[]  & List of headers of uncle's  blocks. \\ \hline
    \end{tabular}

\end{table}

\begin{table}[h!]
    \centering
    \caption{Structure of an Ethereum transaction~\cite{4Wood2014}.}
    \label{tab:4}
    \begin{tabular}{p{1.5cm}p{6.1cm}}\hline
       \textbf{Field Name}  &  \textbf{Description} \\\hline\hline
       once          &  A scalar value equal to the number of transactions sent by the sender.\\
       gasPrice      &  A scalar value equal to the number of Wei to be paid per unit of gas for all computational costs
                        incurred as a result of the execution of the transaction.\\ 
       gasLimit      &  A scalar value equal to the maximum amount of gas that should be used in executing  this  
                        transaction. This is paid up-front, -front, before any computation is done and may not be increased later.\\
       to            &  The 160-bit address of the message call's recipient or,  the zero address for a contract creation transaction.\\
       value         & : A scalar value equal to the number of Wei to  be transferred to the message call’s recipient or,
                        in the case of contract creation, as an endowment to the newly created account. \\
        v, r, s      & Values corresponding to the signature of the  transaction and used to determine the sender of the transaction.  
       \\\hline
    \end{tabular}
\end{table}
 
\begin{table}[h!]
    \centering
      \caption{Ethereum creation contract additional properties .~\cite{4Wood2014}.}
    \label{tab:5}
    \begin{tabular}{p{2.3cm}p{5.3cm}}\hline
       \textbf{Field Name}  &  \textbf{Description} \\\hline\hline
       init & An unlimited size byte array specifying the EVM-code for the account initialisation procedure. 
      \\ \hline
     \end{tabular}
  
\end{table}

\begin{table}[h!]
    \centering
     \caption{Ethereum call message additional properties~\cite{4Wood2014}.}
    \label{tab:6}
    \begin{tabular}{p{2.3cm}p{5.3cm}}\hline
       \textbf{Field Name}  &  \textbf{Description} \\\hline\hline
       data  &  An unlimited size byte array specifying the input data of the message call.
      \\ \hline
     \end{tabular}
   
\end{table}

\begin{table}[h!]
    \centering
      \caption{Structure of a Hyperledger Fabric block~~\cite{5}.}
    \label{tab:7}
    \begin{tabular}{p{2.3cm}p{5.3cm}}\hline
       \textbf{Field Name}  &  \textbf{Description} \\\hline\hline
       number  & An integer that starts at 0, and increases by 1 for each new block added to the chain. \\
       currentBlockHash & The hash of all transactions contained in the block.\\
       previousHash & A copy of the hash of the previous block of the chain.\\
       dataHash & The hash of the data object. \\
       
      \\ \hline
     \end{tabular}
  
\end{table}

\begin{table}[h!]
    \centering
    \caption{Structure of a Hyperledger Fabric transaction~\cite{5}.}
    \label{tab:8}
    \begin{tabular}{p{2.3cm}p{5.3cm}}\hline
       \textbf{Field Name}  &  \textbf{Description} \\\hline\hline
       type & Transaction Type. \\
       version &  Number indicating the current version.\\
       timestamp & Time in which the transaction was created. \\ 
       channelId & Chain with channel identification. \\
       txId & Object with the transaction id.  \\
       epoch &  Time to which the transaction belongs. \\
       payloadVisibility & Payload visibility.  \\
       chaincodePath & Chaincode Route \\
       chaincodeName &  Chain with the name of the chaincode. \\
       chaincodeVersion &  Chaincode version string , example: v1
      \\ \hline
     \end{tabular}

\end{table}

In addition, the existence of classes and properties that answer the competency questions must be guaranteed. For example, for CQ3 question, the ``totalTransactions'' property must be included in the ``Block'' class.

All classes, properties and cardinalities identified for the domain and scope proposed are summarized in the extended entity-relationship diagram of Figure~\ref{fig:6}. This model is widely used in database design and allows incorporating semantic information about the real world \cite{PPSChen}. The type of value to be used in the properties for facets are strings and decimals. 

It is easy to notice the hierarchy of classes and subclasses. For example, BitcoinBlock is a subclass of Block. Also the ``Has'' property between classes must be implemented as object properties.

\subsection{Implementation} 

We had used Protege ontology development tool for implementation in its version 5.5.0. That tool was developed by Stanford University (\url{https://protege.stanford.edu/}). It has a graphic interface with configurable views and tabs that makes it easy to navigate through classes, properties and facets as seen in Figure~\ref{fig:5}.

Disjoint classes are established with the ``DisjointWith'' property. For example EthereumBlock is disjoint with BitcoinBlock and with HyperledgerBlock.

\subsection{INSTANCES}
As a final step we have the creation of instances. For example, with BLONDiE we can generate instances in RDF format describing different concepts of Bitcoin blocks and transactions. To obtain such information, an easy way to explore Bitcoin blockchain is using a tool known as Block Explorer, such as \url{https://blockexplorer.com/}.

We extracted some fields from the block number 1 and from the transaction number 1 from the official Bitcoin chain, then we  mapped them to existing properties in our ontology and presented them in the Figure~\ref{fig:3} and~\ref{fig:4}.

\section{Results and Discussion} 
The result is the ontology in owl format available in the github repository: \url{https://github.com/hedugaro/Blondie}

The ontology has 23 classes, 11 object properties and 64 data properties.

\begin{figure}
    \centering
    \includegraphics[scale = 0.33]{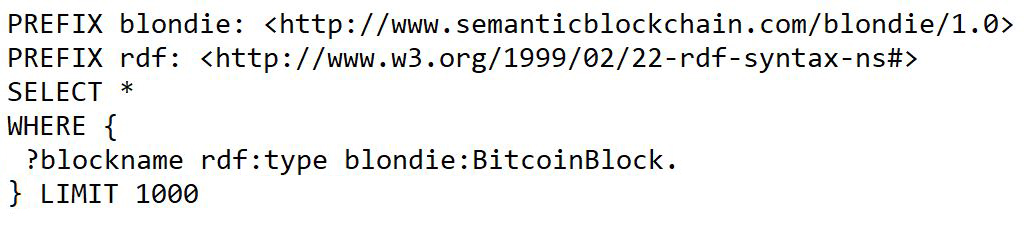}
    \caption{SPARQL query using BLONDiE}
    \label{fig:7}
\end{figure}

The dominant standard for queries on the semantic web is SPARQL. We present a basic query as an example that results in triplets using BLONDiE in Figure~\ref{fig:7}.

Third and Domingue~\cite{third2017linkchains,third2017linked} consider BLONDiE as ``the most developed vocabulary for representing blockchain concepts, with the most potential to enable reusable modelling across different distributed ledgers in the future''. They implemented a semantic index to the main Ethereum blockchain. They exposed blockchain data (blocks and transactions) as Linked Data using BLONDiE ontology to map smart contracts to the Minimal Service Model ontology~\cite{third2017linked}.

Also, BLONDiE is being used as part of a tamper-proof audit solution proposed by Sutton and Samavi~\cite{sutton2017blockchain}.  They used it mainly to generate triplets of Bitcoin transactions. 

The result of the query of Figure~\ref{fig:7} lists the first thousand blocks of Bitcoin.

\section{CONCLUSIONS}

This document summarizes the implementation of an unprecedented ontology for the description of the native structures and related information of the three most relevant blockchain projects to date: Bitcoin, Ethereum and Hyperledger Fabric.

We had used relevant documentation for the extraction of the required elements for the vocabulary and Protege for its implementation, a leading tool in the development of ontologies. The result is BLONDiE in its version 1.0, an ontology easily extensible to other projects.

As our future work we are planning to extend BLONDiE definitions by covering more blockchain projects. Also, we are planning to develope a semantic block explorer capable of querying on SPARQL.

\bibliographystyle{IEEEtran} 
\bibliography{bibtex-ieee}

\begin{thebibliography}{10}
\providecommand{\url}[1]{#1}
\csname url@samestyle\endcsname
\providecommand{\newblock}{\relax}
\providecommand{\bibinfo}[2]{#2}
\providecommand{\BIBentrySTDinterwordspacing}{\spaceskip=0pt\relax}
\providecommand{\BIBentryALTinterwordstretchfactor}{4}
\providecommand{\BIBentryALTinterwordspacing}{\spaceskip=\fontdimen2\font plus
\BIBentryALTinterwordstretchfactor\fontdimen3\font minus
  \fontdimen4\font\relax}
\providecommand{\BIBforeignlanguage}[2]{{%
\expandafter\ifx\csname l@#1\endcsname\relax
\typeout{** WARNING: IEEEtran.bst: No hyphenation pattern has been}%
\typeout{** loaded for the language `#1'. Using the pattern for}%
\typeout{** the default language instead.}%
\else
\language=\csname l@#1\endcsname
\fi
#2}}
\providecommand{\BIBdecl}{\relax}
\BIBdecl

\bibitem{1BHLM2015}
J.~Busse, B.~G. Humm, C.~L{\"u}bbert, F.~Moelter, A.~Reibold, M.~Rewald,
  V.~Schl{\"u}ter, B.~Seiler, E.~Tegtmeier, and T.~Zeh, ``Actually, what does
  “ontology” mean?'' \emph{Journal of computing and information
  technology}, vol.~23, no.~1, pp. 29--41, 2015.

\bibitem{2w3c}
M.~Needleman, ``The w3c semantic web activity,'' \emph{Serials Review},
  vol.~29, no.~1, pp. 63--64, 2003.

\bibitem{3Nakamoto2008}
S.~Nakamoto and A.~Bitcoin, ``A peer-to-peer electronic cash system,''
  \emph{Bitcoin.--URL: https://bitcoin.org/bitcoin.pdf}, 2008.

\bibitem{4Wood2014}
G.~Wood \emph{et~al.}, ``Ethereum: A secure decentralised generalised
  transaction ledger,'' \emph{Ethereum project yellow paper}, vol. 151, no.
  2014, pp. 1--32, 2014.

\bibitem{5}
E.~Androulaki, A.~Barger, V.~Bortnikov, C.~Cachin, K.~Christidis, A.~De~Caro,
  D.~Enyeart, C.~Ferris, G.~Laventman, Y.~Manevich \emph{et~al.}, ``Hyperledger
  fabric: a distributed operating system for permissioned blockchains,'' Tech.
  Rep., 2018.

\bibitem{berners2000semantic}
T.~Berners-Lee, ``Semantic web on xml,''
  \emph{https://www.w3.org/2000/Talks/1206-xml2k-tbl/}, 2000.

\bibitem{7Swanson2014}
T.~Swanson, ``Great chain of numbers: A guide to smart contracts, smart
  property and trustless asset management,'' \emph{Amazon Digital Services},
  2014.

\bibitem{8Szabo1997}
N.~Szabo, ``Formalizing and securing relationships on public networks,''
  \emph{First Monday}, vol.~2, no.~9, 1997.

\bibitem{9Hector2016}
H.~Ugarte, ``A more pragmatic web 3.0: Linked blockchain data,'' \emph{Bonn,
  Germany}, 2017.

\bibitem{10NDMc2001}
N.~F. Noy, D.~L. McGuinness \emph{et~al.}, ``Ontology development 101: A guide
  to creating your first ontology,'' 2001.

\bibitem{Antono2014}
A.~M. Antonopoulos, \emph{Mastering Bitcoin: unlocking digital
  cryptocurrencies}.\hskip 1em plus 0.5em minus 0.4em\relax " O'Reilly Media,
  Inc.", 2014.

\bibitem{bitcoin2015}
\BIBentryALTinterwordspacing
Bitcoin.org, ``Bitcoin developer guide, find detailed information about the
  bitcoin protocol and related specifications.'' \emph{Bitcoin.org}, 2015.
  [Online]. Available:
  \url{http://docshare04.docshare.tips/files/27195/271951119.pdf}
\BIBentrySTDinterwordspacing

\bibitem{miller1998introduction}
E.~Miller, ``An introduction to the resource description framework,''
  \emph{Bulletin of the American Society for Information Science and
  Technology}, vol.~25, no.~1, pp. 15--19, 1998.

\bibitem{Gutierrez2005}
C.~Gutierrez, C.~Hurtado, and A.~Vaisman, ``Temporal rdf,'' in \emph{European
  Semantic Web Conference}.\hskip 1em plus 0.5em minus 0.4em\relax Springer,
  2005, pp. 93--107.

\bibitem{14Cardoso2015}
J.~Cardoso and A.~M. Pinto, ``The web ontology language (owl) and its
  applications,'' in \emph{Encyclopedia of Information Science and Technology,
  Third Edition}.\hskip 1em plus 0.5em minus 0.4em\relax IGI Global, 2015, pp.
  7662--7673.

\bibitem{Towards2017}
J.~De~Kruijff and H.~Weigand, ``Towards a blockchain ontology,'' \emph{The
  Netherlands, pdfs. semanticscholar.
  org/0782/c5badb4f407ee0964d07eda9f74a92de3298. pdf [dost{\d{e}}p 22.07.
  2018]}, 2017.

\bibitem{okupski2014bitcoin}
K.~Okupski, ``Bitcoin developer reference,'' in \emph{Eindhoven}, 2014.

\bibitem{PPSChen}
P.~P.-S. Chen, ``The entity-relationship model—toward a unified view of
  data,'' \emph{ACM transactions on database systems (TODS)}, vol.~1, no.~1,
  pp. 9--36, 1976.

\bibitem{third2017linkchains}
A.~Third and J.~Domingue, ``Linkchains: Exploring the space of decentralised
  trustworthy linked data,'' 2017.

\bibitem{third2017linked}
A.~\vspace{0mm}Third and J.~Domingue, ``Linked data indexing of distributed
  ledgers,'' pp. 1431--1436, 2017.

\bibitem{sutton2017blockchain}
A.~Sutton and R.~Samavi, ``Blockchain enabled privacy audit logs,'' in
  \emph{International Semantic Web Conference}.\hskip 1em plus 0.5em minus
  0.4em\relax Springer, 2017, pp. 645--660.

\end{thebibliography}

\end{document}